\documentstyle[12pt,twoside]{article}

\def\a{\alpha}
\def\b{\beta}

\def\d{\delta}
\def\f{\phi}                    

\def\h{\eta}

\def\k{\kappa}
\def\l{\lambda}
\def\m{\mu}
\def\n{\nu}
\def\o{\omega}
\def\p{\pi}                     
\def\r{\rho}                    
\def\s{\sigma}                  

\def\x{\xi}

\def\D{\Delta}

\def\L{\Lambda}

\def\cf{{\cal F}}

\def\ch{{\cal H}}

\catcode`@=11

\def\un#1{\relax\ifmmode\@@underline#1\else $\@@underline{\hbox{#1}}$\relax\fi}

{\catcode`\'=\active \gdef'{{}~\bgroup\prim@s}}

\def\magstep#1{\ifcase#1 \@m\or 1200\or 1440\or 1728\or 2074\or 2488\or
        2986\fi\relax}

\font\twfvmi=cmmi10\@magscale5
    \skewchar\twfvmi='177
\font\twfvsy=cmsy10\@magscale5
    \skewchar\twfvsy='60
\font\twfvly=lasy10\@magscale5
\font\thtyrm=cmr10\@magscale6

\def\vpt{\textfont\z@\fivrm
  \scriptfont\z@\fivrm \scriptscriptfont\z@\fivrm
\textfont\@ne\fivmi \scriptfont\@ne\fivmi \scriptscriptfont\@ne\fivmi
\textfont\tw@\fivsy \scriptfont\tw@\fivsy \scriptscriptfont\tw@\fivsy
\textfont\thr@@\tenex \scriptfont\thr@@\tenex \scriptscriptfont\thr@@\tenex
\def\prm{\fam\z@\fivrm}%
\def\unboldmath{\everymath{}\everydisplay{}\@nomath
  \unboldmath\fam\@ne\@boldfalse}\@boldfalse
\def\boldmath{\@subfont\boldmath\unboldmath}%
\def\pit{\@getfont\pit\itfam\@vpt{cmti5}}%
\def\psl{\@subfont\sl\it}%
\def\pbf{\@getfont\pbf\bffam\@vpt{cmbx5}}%
\def\ptt{\@subfont\tt\rm}%
\def\psf{\@subfont\sf\rm}%
\def\psc{\@subfont\sc\rm}%
\def\ly{\fam\lyfam\fivly}\textfont\lyfam\fivly
    \scriptfont\lyfam\fivly \scriptscriptfont\lyfam\fivly
\@setstrut\rm}

\def\@vpt{}

\def\vipt{\textfont\z@\sixrm
  \scriptfont\z@\sixrm \scriptscriptfont\z@\sixrm
\textfont\@ne\sixmi \scriptfont\@ne\sixmi \scriptscriptfont\@ne\sixmi
\textfont\tw@\sixsy \scriptfont\tw@\sixsy \scriptscriptfont\tw@\sixsy
\textfont\thr@@\tenex \scriptfont\thr@@\tenex \scriptscriptfont\thr@@\tenex
\def\prm{\fam\z@\sixrm}%
\def\unboldmath{\everymath{}\everydisplay{}\@nomath
  \unboldmath\@boldfalse}\@boldfalse
\def\boldmath{\@subfont\boldmath\unboldmath}%
\def\pit{\@subfont\it\rm}%
\def\psl{\@subfont\sl\rm}%
\def\pbf{\@getfont\pbf\bffam\@vipt{cmbx6}}%
\def\ptt{\@subfont\tt\rm}%
\def\psf{\@subfont\sf\rm}%
\def\psc{\@subfont\sc\rm}%
\def\ly{\fam\lyfam\sixly}\textfont\lyfam\sixly
    \scriptfont\lyfam\sixly \scriptscriptfont\lyfam\sixly
\@setstrut\rm}

\def\@vipt{}

\def\xxxpt{\textfont\z@\thtyrm
  \scriptfont\z@\twfvrm \scriptscriptfont\z@\twtyrm
\textfont\@ne\twfvmi \scriptfont\@ne\twfvmi \scriptscriptfont\@ne\twtymi
\textfont\tw@\twfvsy \scriptfont\tw@\twfvsy \scriptscriptfont\tw@\twtysy
\textfont\thr@@\tenex \scriptfont\thr@@\tenex \scriptscriptfont\thr@@\tenex
\def\unboldmath{\everymath{}\everydisplay{}\@nomath\unboldmath
        \textfont\@ne\twfvmi \textfont\tw@\twfvsy \textfont\lyfam\twfvly
        \@boldfalse}\@boldfalse
\def\boldmath{\@subfont\boldmath\unboldmath}%
\def\prm{\fam\z@\thtyrm}%
\def\pit{\@subfont\it\rm}%
\def\psl{\@subfont\sl\rm}%
\def\pbf{\@getfont\pbf\bffam\@xxxpt{cmbx10\@magscale6}}%
\def\ptt{\@subfont\tt\rm}%
\def\psf{\@subfont\sf\rm}%
\def\psc{\@subfont\sc\rm}%
\def\ly{\fam\lyfam\twfvly}\textfont\lyfam\twfvly
   \scriptfont\lyfam\twfvly \scriptscriptfont\lyfam\twtyly
\@setstrut \rm}

\def\@xxxpt{}

\def\Huge{\@setsize\Huge{36pt}\xxxpt\@xxxpt}

\font\thtymi=cmmi10\@magscale6
    \skewchar\thtymi='177
\font\thtysy=cmsy10\@magscale6
    \skewchar\thtysy='60
\font\thtyly=lasy10\@magscale6
\font\thsirm=cmr12\@magscale6

\def\xxxvipt{\textfont\z@\thsirm
  \scriptfont\z@\thtyrm \scriptscriptfont\z@\twfvrm
\textfont\@ne\thtymi \scriptfont\@ne\thtymi \scriptscriptfont\@ne\twfvmi
\textfont\tw@\thtysy \scriptfont\tw@\thtysy \scriptscriptfont\tw@\twfvsy
\textfont\thr@@\tenex \scriptfont\thr@@\tenex \scriptscriptfont\thr@@\tenex
\def\unboldmath{\everymath{}\everydisplay{}\@nomath\unboldmath
        \textfont\@ne\thtymi \textfont\tw@\thtysy \textfont\lyfam\thtyly
        \@boldfalse}\@boldfalse
\def\boldmath{\@subfont\boldmath\unboldmath}%
\def\prm{\fam\z@\thsirm}%
\def\pit{\@subfont\it\rm}%
\def\psl{\@subfont\sl\rm}%
\def\pbf{\@getfont\pbf\bffam\@xxxpt{cmss12\@magscale6}}%
\def\ptt{\@subfont\tt\rm}%
\def\psf{\@subfont\sf\rm}%
\def\psc{\@subfont\sc\rm}%
\def\ly{\fam\lyfam\thtyly}\textfont\lyfam\thtyly
   \scriptfont\lyfam\thtyly \scriptscriptfont\lyfam\twfvly
\@setstrut \rm}

\def\@xxxvipt{}

\def\HUGE{\@setsize\HUGE{43pt}\xxxvipt\@xxxvipt}

\catcode`@=12

\font\tenex=cmex10 scaled 1200

\def\Sc#1{\hbox{\sc #1}}        

\def\bo{{\raise.05ex\hbox{\large$\Box$}\:}}             
\def\cbo{{\,\raise-.15ex\Sc [\,}}                       
\def\pa{\partial}                                       
\def\su{\sum}                                           
\def\TH{{\raise.2ex\hbox{$\displaystyle \bigodot$}\mskip-4.7mu \llap H \;}}
\def\face{\hbox{\normalsize$\;\;\:{\raise.9ex\hbox{\oo n}\mskip-13mu \llap
        {${\buildrel{\hbox{\frtnrm ..}}\over\smile}$}}\:$}}     
\def\Face{{\raise.2ex\hbox{$\displaystyle \bigodot$}\mskip-2.2mu \llap {$\ddot
        \smile$}}}                                      
\def\Lhat{{\bf\rlap{\kern-.09em$\hat{\phantom L}$}L}}
\def\Lcheck{{\bf\rlap{\kern-.09em$\check{\phantom L}$}L}}

\def\sp#1{{}^{#1}}                              
\def\sb#1{{}_{#1}}                              
\def\sl#1{\rlap{\hbox{$\mskip 1 mu /$}}#1}      
\def\sbra#1{\left\langle #1\right|}             
\def\sket#1{\left| #1\right\rangle}             
\def\svev#1{\left\langle #1\right\rangle}       
\def\leftrightarrowfill{$\mathsurround=0pt \mathord\leftarrow \mkern-6mu
        \cleaders\hbox{$\mkern-2mu \mathord- \mkern-2mu$}\hfill
        \mkern-6mu \mathord\rightarrow$}
\def\dvec#1{\vbox{\ialign{##\crcr
        \leftrightarrowfill\crcr\noalign{\kern-1pt\nointerlineskip}
        $\hfil\displaystyle{#1}\hfil$\crcr}}}           
\def\dt#1{{\buildrel {\hbox{\LARGE .}} \over {#1}}}     
\def\ddt#1{{\buildrel {\hbox{\LARGE .\kern-2pt.}} \over {#1}}}

\def\frac#1#2{{\textstyle{#1\over\vphantom2\smash{\raise.20ex
        \hbox{$\scriptstyle{#2}$}}}}}                   
\def\ha{\frac12}                                        
\def\sfrac#1#2{{\vphantom1\smash{\lower.5ex\hbox{\small$#1$}}\over
        \vphantom1\smash{\raise.4ex\hbox{\small$#2$}}}} 
\def\bfrac#1#2{{\vphantom1\smash{\lower.5ex\hbox{$#1$}}\over
        \vphantom1\smash{\raise.3ex\hbox{$#2$}}}}       
\def\afrac#1#2{{\vphantom1\smash{\lower.5ex\hbox{$#1$}}\over#2}}    

\def\boxes#1{
        \newcount\num
        \num=1
        \newdimen\downsy
        \downsy=-1.64ex
        \mskip-7.8mu
        \bo
        \loop
        \ifnum\num<#1
        \llap{\raise\num\downsy\hbox{$\bo$}}
        \advance\num by1
        \repeat}
\def\boxup#1#2{\newcount\numup
        \numup=#1
        \advance\numup by-1
        \newdimen\upsy
        \upsy=.82ex
        \mskip7.8mu
        \raise\numup\upsy\hbox{$#2$}}

\newskip\humongous \humongous=0pt plus 1000pt minus 1000pt
\def\caja{\mathsurround=0pt}

\newif\ifdtup
\def\panorama{\global\dtuptrue \openup2\jot \caja
        \everycr{\noalign{\ifdtup \global\dtupfalse
        \vskip-\lineskiplimit \vskip\normallineskiplimit
        \else \penalty\interdisplaylinepenalty \fi}}}
\def\li#1{\panorama \tabskip=\humongous                         
        \halign to\displaywidth{\hfil$\displaystyle{##}$
        \tabskip=0pt&$\displaystyle{{}##}$\hfil
        \tabskip=\humongous&\llap{$##$}\tabskip=0pt
        \crcr#1\crcr}}

\def\CMP{Commun. Math. Phys.}

\def\PL{Phys. Lett. }

\def\PRD{Phys. Rev. D}
\def\ref#1{$\sp{#1]}$}

\parskip=\medskipamount                 
\def\baselinestretch{1.2}       
\thicklines                         
\def\title#1#2#3#4{
\begin{document}
        {\hbox to\hsize{#4 \hfill  #3}}\par
        \begin{center}\vskip.5in minus.1in {\Large\bf #1}\\[.5in minus.2in]{#2}
        \vskip1.4in minus1.2in {\bf ABSTRACT}\\[.1in]\end{center}
        \begin{quotation}\par}
\def\author#1#2{#1\\[.1in]{\it #2}\\[.1in]}

\def\AMIC{Aleksandar Mikovic\'c
\\[.1in]{\it Blackett Laboratory, Imperial College, Prince Consort Road, London
SW7 2BZ, UK}\\[.1in]}

\def\AMICIF{Aleksandar Mikovi\'c\,
\footnote{Work supported by MNTRS and Royal Society}
\\[.1in] {\it Blackett Laboratory, Imperial College, Prince Consort
Road, London SW7 2BZ, UK}\\[.1in]
and \\[.1 in]
{\it Institute of Physics, P.O. Box 57, 11001 Belgrade, Yugoslavia}
\footnote{Permanent address}\\ {\it E-mail:\, mikovic@castor.phy.bg.ac.yu}}

\def\AMSISSA{Aleksandar Mikovi\'c\,
\footnote{E-mail address: mikovic@castor.phy.bg.ac.yu}
\\[.1in] {\it SISSA-International School for Advanced Studies\\
Via Beirut 2-4, Trieste 34100, Italy}\\[.1in]
and \\[.1 in]
{\it Institute of Physics, P.O. Box 57, 11001 Belgrade, Yugoslavia}
\footnote{Permanent address}}

\def\AM{Aleksandar Mikovi\'c
\footnote{E-mail address: mikovic@castor.phy.bg.ac.yu}
\\[.1in] {\it Institute of Physics, P.O.Box 57, Belgrade 11001, Yugoslavia}
\\[.1in]}

\def\AMsazda{Aleksandar Mikovi\'c
\footnote{E-mail address: mikovic@castor.phy.bg.ac.yu}
and Branislav Sazdovi\'c \footnote{E-mail: sazdovic@castor.phy.bg.ac.yu}
\\[.1in] {\it Institute of Physics, P.O.Box 57, Belgrade 11001, Yugoslavia}
\\[.1in]}

\def\endtitle{\par\end{quotation}\vskip3.5in minus2.3in\newpage}


\def\endabstract{\par\end{quotation}
        \renewcommand{\baselinestretch}{1.2}\small\normalsize}


\def\xpar{\par}                                         

\def\letterhead{
        \centerline{\large\sf INSTITUTE OF PHYSICS}
        \centerline{\sf P.O.Box 57, 11001 Belgrade, Yugoslavia}
        \rightline{\scriptsize\sf Dr Aleksandar Mikovi\'c}
        \vskip-.07in
        \rightline{\scriptsize\sf Tel: 11 615 575}
        \vskip-.07in
        \rightline{\scriptsize\sf E-mail: MIKOVIC@CASTOR.PHY.BG.AC.YU}}

\def\sig#1{{\leftskip=3.75in\parindent=0in\goodbreak\bigskip{Sincerely yours,}
\nobreak\vskip .7in{#1}\par}}

\def\ssig#1{{\leftskip=3.75in\parindent=0in\goodbreak\bigskip{}
\nobreak\vskip .7in{#1}\par}}


\def\ree#1#2#3{
        \hfuzz=35pt\hsize=5.5in\textwidth=5.5in
        \begin{document}
        \ttraggedright
        \par
        \noindent Referee report on Manuscript \##1\\
        Title: #2\\
        Authors: #3}


\def\start#1{\pagestyle{myheadings}\begin{document}\thispagestyle{myheadings}
        \setcounter{page}{#1}}


\catcode`@=11

\def\ps@myheadings{\def\@oddhead{\hbox{}\footnotesize\bf\rightmark \hfil
        \thepage}\def\@oddfoot{}\def\@evenhead{\footnotesize\bf
        \thepage\hfil\leftmark\hbox{}}\def\@evenfoot{}
        \def\sectionmark##1{}\def\subsectionmark##1{}
        \topmargin=-.35in\headheight=.17in\headsep=.35in}
\def\ps@acidheadings{\def\@oddhead{\hbox{}\rightmark\hbox{}}
        \def\@oddfoot{\rm\hfil\thepage\hfil}
        \def\@evenhead{\hbox{}\leftmark\hbox{}}\let\@evenfoot\@oddfoot
        \def\sectionmark##1{}\def\subsectionmark##1{}
        \topmargin=-.35in\headheight=.17in\headsep=.35in}

\catcode`@=12

\def\sect#1{\bigskip\medskip\goodbreak\noindent{\large\bf{#1}}\par\nobreak
        \medskip\markright{#1}}
\def\chsc#1#2{\phantom m\vskip.5in\noindent{\LARGE\bf{#1}}\par\vskip.75in
        \noindent{\large\bf{#2}}\par\medskip\markboth{#1}{#2}}
\def\Chsc#1#2#3#4{\phantom m\vskip.5in\noindent\halign{\LARGE\bf##&
        \LARGE\bf##\hfil\cr{#1}&{#2}\cr\noalign{\vskip8pt}&{#3}\cr}\par\vskip
        .75in\noindent{\large\bf{#4}}\par\medskip\markboth{{#1}{#2}{#3}}{#4}}
\def\chap#1{\phantom m\vskip.5in\noindent{\LARGE\bf{#1}}\par\vskip.75in
        \markboth{#1}{#1}}
\def\refs{\bigskip\medskip\goodbreak\noindent{\large\bf{REFERENCES}}\par
        \nobreak\bigskip\markboth{REFERENCES}{REFERENCES}
        \frenchspacing \parskip=0pt \renewcommand{\baselinestretch}{1}\small}
\def\unrefs{\normalsize \nonfrenchspacing \parskip=medskipamount}
\def\Item{\par\hang\textindent}
\def\Itemitem{\par\indent \hangindent2\parindent \textindent}
\def\makelabel#1{\hfil #1}
\def\topic{\par\noindent \hangafter1 \hangindent20pt}
\def\Topic{\par\noindent \hangafter1 \hangindent60pt}

\title{Unitary Theory of Evaporating 2D Black Holes}
{\AMICIF}{Imperial/TP/94-95/50}{July 1995}

We study a manifestly unitary formulation of 2d dilaton quantum
gravity based on the reduced phase space quantization.
The spacetime metric operator
can be expanded in a formal power series of the matter energy-momentum
tensor operator. This expansion can be used for calculating the quantum
corrections to the classical black hole metric by evaluating the expectation
value of the metric operator in an appropriate class of the physical states.
When the normal ordering in the metric operator
is chosen to be with respect to Kruskal vacuum, the lowest order
semiclassical metric is exactly the one-loop effective action metric
discovered by Bose, Parker and Peleg. The corresponding semiclassical geometry
describes an evaporating black hole which ends
up as a remnant. The calculation of higher order corrections and implications
for the black hole fate are discussed.

\endtitle

\sect{1. Introduction}

Hawking's discovery that black holes radiate thermally \cite{hawk},
gave rise to a series of questions about the quantum fate of a black
hole. Is the quantum evolution of an evaporating black hole unitary or not?
Related to that are the questions of information loss and the final
spacetime geometry. A
proper framework for answering these questions would be a quantum theory of
gravity, but in the absence of such a theory the best one can do is to
study toy models of black hole formation and evaporation.

Two-dimensional dilaton gravity models coupled to matter (CGHS and its
modifications \cite{{cghs},{rev},{rst}}) represent such toy models, and
they have been extensively studied due to their classical and quantum
solvabilty. One aspect of the quantum solvability is the fact that the
exact canonical constrained quantization can be performed for $N=24$
matter fields and the physical Hilbert space can be
obtained, which is the free-field matter Fock space
\cite{{mik1},{hks},{mik2},{mik3}}.
Furthermore, it
has been noticed that the constraints can be deparametrized, or
equivalently a time variable can be found \cite{mik2,mik3}, which
allows for a construction of a unitary quantum evolution operator. This
can be done explicitely in the reduced phase space formalism (i.e. by
fixing a gauge and solving the constraints in terms of independent canonical
variables), and a unitary quantum theory can be constructed
\cite{mik4}. The metric operator in the Heisenberg picture can be
expressed as an operator valued solution of the classical equations of
motion, and its
expectation value in a physical state determines an effective
quantum metric.

Calculating the quantum corrections and backreaction turns out to be
easier in this
unitary gauge formalism then in the effective action approach.
One expands the metric
operator into a formal power series in the matter energy-momentum tensor
operator and takes the expectation value in the appropriate physical
state \cite{mik4}.
By choosing a normal ordering in the metric operator to be
with respect to the ``out'' vacuum corresponding to the background
of the classical black
hole solution, and by taking the initial state to be a coherent state
with respect to the
``in'' dilaton vacuum, the lowest order semiclassical metric was obtained.
It was the black hole metric plus the back-reaction correction due
to the Hawking radiation. This metric was different from the other
one-loop semiclassical metrics obtained in the effective
action approach, and it was
conjectured that different normal orderings in the unitary gauge
formalism correspond to
different combinations of the one-loop counterterms in the effective action.

In this paper we show that a normal ordering in the metric operator
with respect to Kruskal vacuum gives the lowest order
semiclassical metric  which is the same as the semiclassical metric
obtained by Bose, Parker and Peleg (BPP) from a one-loop effective action
\cite{bpp}. In section 2 we describe the reduced phase space
formalism for the CGHS
model. In section 3 we quantize the reduced phase space
formulation, and define the
quantum metric and the class of relevant physical states. In section 4
we show how the BPP solution arises in the semiclassical limit of our
quantum theory. In section 5 we describe how to calculate the second order
quantum corrections, and in section 6 we present our
conclussions.

\sect{2. Reduced phase space formalism}

In the reduced phase space formalism one solves the constraints
classicaly and then quantize the independent set of canonical
variables. Solving the constraints requires a gauge fixing, and one
has to take care that a good gauge is chosen. The advantage of this
method is that one can obtain the physical Hilbert space relatively easy.
The disadvantage is that the gauge symmetries (which are  generated
by the constraints) are not manifest. In the case of black hole toy
models, this strategy was used for analysing the spherically symmetric
black hole collaps (BCMN model) \cite{{bcmn},{unru},{haj}},
which can be recasted in the 2d dilaton gravity form \cite{rev}. However,
the problem there is
that the BCMN gauge does not penetrate the horizon, and consequently
the Hamiltonian evolution stops at the horizon \cite{haj}. Also the
Hamiltonian of the BCMN model is a non-local function of the matter
fields, so it is difficult to promote it into a Hermitian
operator. In the case of the CGHS model such problems are absent
\cite{mik4}, and the reduced phase space approach can be implemented.

We start from the classical CGHS action \cite{cghs}
$$ S =  \int_{M} d^2 x \sqrt{-g} \left[ e^{-\f}\left( R +
 (\nabla \f)^2 + 4\l^2 \right) - \ha\su_{i=1}^N (\nabla f_i)^2 \right]\quad,
\eqno(2.1)$$
where $\f$ is a dilaton, $f_i$ are scalar matter fields,  $g$, $R$
and $\nabla$ are
determinant, scalar curvature and covariant derivative respectively,
associated with a metric $g_{\m\n}$ on the 2d manifold $M$. Topology of
$M$ is that of $ {\bf R} \times {\bf R}$. We make a field redefinition
$$ \tilde{g}_{\m\n} = e^{-\f} g_{\m\n} \quad,\quad \tilde{\f} = e^{-\f} \quad,
\eqno(2.2)$$
so that one obtains a simpler action
$$ S =  \int_{M} d^2 x \sqrt{-\tilde{g}}\left[
\tilde{R}\tilde{\f} + 4\l^2 -\ha\su_{i=1}^N (\tilde{\nabla} f_i)^2 \right]
\quad. \eqno(2.3)$$
Canonical analysis of the action (2.3) gives
$$ S= \int dt dx \left( \p_{\tilde\r}\dt{\tilde\r} +
\p_{\tilde\f}\dt{\tilde\f} + \p_{f}\dt{f} -
 N_0 G_0 - N_1 G\sb 1 \right) - \int dt H \quad, \eqno(2.4)$$
where $N_0$ and $N_1$ are the rescaled laps and shift, while
the Hamiltonian constraint $G_0$ and the spatial diffeomorphisms
constraint $G_1$ are given as
$$\li{G\sb 0  =& - \p_{\tilde\r} \p_{\tilde\f}  - 4\l^2 e^{\tilde\r} +
2\tilde{\f}^{\prime\prime} - \tilde{\r}^{\prime}\tilde{\f}^{\prime} +
 \ha (\p_f^2 + (f^{\prime})^2)\cr
G\sb 1  =& \p_{\tilde{\f}} {\tilde{\f}}^{\prime} + \p_{\tilde{\r}}
\tilde{\r}^{\prime} -2 \p_{\tilde{\r}}^{\prime}
+ \p_f f^{\prime} \quad.&(2.5)\cr} $$
The primes stand for the $x$ derivatives, $\tilde\r$
is the conformal factor ($\tilde{g} = e^{\tilde\r}$), and we have
taken a single matter field ($N=1$)
for the simplycity sake. The boundary term in (2.4) is needed in order
to obtain correct equations of motion \cite{bk}, and $H$ will be the
Hamiltonian in the physical gauge.

Now we fix the gauge
$$ \tilde\r = 0 \quad,\quad \p_{\tilde\f }= 0 \quad,\eqno(2.6)$$
which can be thoutght of as the 2d dilaton gravity analog of the BCMN
gauge \cite{bcmn}.
However, unlike the BCMN gauge, (2.6) penetrates the horizon and corresponds
to the classical CGHS solution in the gauge $\r = \f$ \cite{mik4}.
Solving the constraints gives
$$\tilde\f = a + bx + \l^2 x^2 - \frac14 \int dx\int dx (\p_f^2 +
(f^{\prime})^2) \quad,\quad \p_{\tilde\r} = c + \ha\int dx \p_f f^{\prime}
\quad, \eqno(2.7)$$
so that the independent canonical variables (or true degrees of freedom) are
$(\p_f, f)$ canonical pairs. The reduced phase space Hamiltonian is a
free-field Hamiltonian
$$ H = \ha\int_{-\infty}^{\infty}dx\,(\p_f^2 +
(f^{\prime})^2)\quad. \eqno(2.8)$$
The dilaton and the original metric can be expressed in the gauge (6) as
$$e^{-\f} =  - \l^2 x^+ x^- - F_+ - F_- \quad,\quad ds^2 = -e^{\f}dx^+ dx^-
\quad,\eqno(2.9)$$
where
$$ F_{\pm}= a_{\pm} + b_{\pm}x^{\pm} +
\int^{x^{\pm}} dy \int^y dz T_{\pm\pm} (z)
\quad.\eqno(2.10)$$
The independent integration constants are $a_+ + a_-$ and $b_{\pm}$,
and $T_{\pm\pm}$ is the matter energy-momentum tensor
$$T_{\pm\pm} = \ha \pa_{\pm} f \pa_{\pm} f \quad.\eqno(2.11)$$
An equivalent form of the solution (2.10), which is suitable for our purposes,
is given by
$$ F_{\pm}=\a_{\pm} + \b_{\pm}x^{\pm} + \int_{\L^{\pm}}^{x^{\pm}} dy
(x^{\pm} - y) T_{\pm\pm} (y)
\quad.\eqno(2.12)$$

The formulas (2.9-10) can be derived from the eq. (2.7) by using
$\p_f = \dt{f}$.

\sect{3. Quantum theory}

Quantum theory is defined by choosing a representation of the quantum
canonical commutation relations
$$ [\p_f (x), f(y) ] = -i \d(x-y) \quad.\eqno(3.1)$$
We take the standard Fock space representation, by defining the creation and
anhilation operators $a^{\dagger},a$ as
$$ a_k = {-i \p_k + k\,{\rm sign}(k) f_k\over\sqrt{2|k|}} \quad,\eqno(3.2)$$
where
$$f(x)=\int_{-\infty}^{\infty}{dk\over\sqrt{2\p}}e^{ikx} f_k \quad,\quad
\p_f (x)=\int_{-\infty}^{\infty}{dk\over\sqrt{2\p}}e^{ikx}\p_k \quad,\eqno(3.3)
$$
so that eq. (3.1) is equivalent to
$$ [a_k , a_q^{\dagger} ] = \d (k-q) \quad.\eqno(3.4)$$
The Fock space ${\cf} (a_k)$ with the vacuum $\sket{0}$ is the physical
Hilbert space of the theory. The
Hamiltonian (2.8) can be promoted into a Hermitian operator acting on ${\cf}$
as
$$H = \int_{-\infty}^{\infty} dk\, \o_k a_{k}^{\dagger} a_k + E_0 \quad
\eqno(3.5)$$
where $\o_k = |k|$ and $E_0$ is the vacuum energy.
Therefore one has a unitary evolution described by a Schr\"odinger
equation
$$ i{\pa\over\pa t} \Psi (t) = H \Psi(t) \quad,\eqno(3.6) $$
where $\Psi(t)$ is a normalisable state from $ \cf$.
It will be convenient to work in the Heisenberg picture
$$ \Psi_0 = e^{iHt}\Psi(t) \quad,\quad A (t) = e^{iHt}Ae^{-iHt}\quad,
\eqno(3.7)$$
so that
$$f(t,x) = e^{iHt} f(x) e^{-iHt} = {1\over\sqrt{2\p}}\int_{-\infty}^{\infty}
{dk\over\sqrt{2\o_k}}\left[ a_k e^{i(kx-\o_k t)} + a_k^{\dagger}
e^{-i(kx-\o_k t)}\right]\quad.\eqno(3.8)$$
It is also useful to split eq. (3.8) into left and right moving parts, so that
$f = f_+ + f_-$ where
$$f_{\pm} (x^{\pm}) = {1\over\sqrt{2\p}}\int_{0}^{\infty}
{dk\over\sqrt{2\o_k}}\left[ a_{\pm}(k) e^{-ikx^{\pm}} + a_{\pm}(k)^{\dagger}
e^{ikx^{\pm}}\right]\quad.\eqno(3.9)$$

The metric is given by the operator $e^\f$, which can be defined as
the inverse of the
$e^{-\f}$ operator. The $e^{-\f}$ operator in the Heisenberg picture
can be easily defined from the expressions (2.9) and (2.10), where now
$f$  is the
operator given by eq. (3.8), while in the expressions for $T_{\pm\pm}$ there
will be a normal ordering with respect to some vacuum in $\cf$, which can be
different from $\sket{0}$, as was the case in \cite{mik4}. However, in
this paper we are going to explore the case of Kruskal vacuum normal ordering.

Given a physical state $\Psi_0$, one can associate an effective metric to
$\Psi (t)$ as
$$e^{\r_{eff} (t,x)} = \sbra{\Psi (t)}e^{\f (x)}\sket{\Psi (t)}
=\sbra{\Psi_0}e^{\f (t,x)}\sket{\Psi_0} \quad.\eqno(3.10)$$
Note that $e^{\r_{eff}}$ can be interpreted as a metric only if the
quantum fluctuations are small. A criterion for this would be
$$\sqrt{ | \svev{e^{2\f}} - \svev{e^\f}^2  |} << \svev{e^\f}\quad.
\eqno(3.11)$$
Note that the dispersion in (3.11) cannot be zero, since otherwise
it would mean that $\Psi_0$ is
an eigenvalue of the metric. This would imply that $\Psi_0$ is not a
normalisable state, since the spectrum of the metric is continious.

In order to calculate $e^{\r_{eff}}$, we use the following formal identity
$$(-\l^2 x^+ x^- -  F )^{-1} = e^{\f_0}(1 - e^{\f_0}\d F)^{-1} =
e^{\f_0} \su_{n=0}^{\infty} e^{n\f_0} \d F^n \quad,\eqno(3.12)$$
where $F_0$ is a c-number function, $e^{-\f_0} = -\l^2 x^+ x^- -  F_0 $ and
$\d F = F - F_0$. Then
$$\svev{(-\l^2 x^+ x^- -  F )^{-1}} =
e^{\f_0} \su_{n=0}^{\infty} e^{n\f_0} \svev{\d F^n}\quad.\eqno(3.13)$$
Note that the expression (3.13) gets slightly simplified if one choses
$$ F_0 = \svev{F_+} + \svev{F_-}\eqno(3.14) $$
since then the $n=1$ term vanishes, and the lowest order metric gives
a one loop semiclassical metric
$$ e^{-\f_0} = -\l^2 x^+ x^- -  \svev{F_+} - \svev{F_-}\quad.\eqno(3.15) $$

We now want to choose $\Psi_0$ such that it is as close as possible to the
classical matter distribution $f_0 (x^+)$ describing  a left-moving pulse of
matter. The corresponding classical metric is described by
$$ e^{-\r} = {M(x^+)\over \l} - \l^2 x^+ \D (x^+) - \l^2 x^+ x^-
\eqno(3.16)$$
where
$$ M (x^+) = \l\int_{-\infty}^{x^+} dy\, y \,T_{++}^0 (y)\quad,\quad
\l^2 \D = \int_{-\infty}^{x^+} dy\, T^0_{++} (y) \quad \eqno(3.17)$$
and $T_{++}^0 = \ha \pa_{+}f_0 \pa_{+} f_0$. The geometry is that of the black
hole of the mass
$$M = \lim_{x^+ \to +\infty} M(x^+)\eqno(3.18)$$
and the horizon is at
$$ x^- = -\D = -\lim_{x^+ \to +\infty} \D (x^+)\quad.\eqno(3.19)$$
The asymptotically
flat coordinates $(\h^+,\h^-)$ at the past null infinity  are given by
\cite{gn}
$$ \l x^+ = e^{\l \h^+} \quad,\quad \l x^- = - e^{-\l \h^-} \quad,\eqno(3.20)$$
while the asymptotically flat coordinates $(\s^+,\s^-)$ at the future
null infinity satisfy
$$ \l x^+ = e^{\l \s^+} \quad,\quad \l (x^- + \D ) = - e^{-\l \s^-}\quad.
\eqno(3.21)$$

Note that a change of coordinates defines a new set of creation and
anhilation operators through
$$f_+ = {1\over\sqrt{2\p}}\int_{0}^{\infty}
{dk\over\sqrt{2\o_k}}\left[ b(k) e^{-ik\h^+} + b(k)^{\dagger}
e^{ik\h^+}\right]\quad,\eqno(3.22)$$
and similarly for the right-moving sector. The old and the new creation
and anhilation operators are related by the Bogoluibov transformations
\cite{gn}. However, one has to keep in mind that in our approach there
is no background metric, so that the coordinates (3.20) and (3.21) could
not be interpreted as the ``in" and the ``out" coordinates.
Only in the regions of the spacetime
where the relation (3.11) is satisfied, one can have a background
geometry. Moreover, this background geometry can be close to the
classical geometry (3.16) for early times \cite{mik4}, so that the
coordinates (3.20) can be still interpreted as the ``in" coordinates.

Hence we take for $\Psi_0$ a coherent state
$$\Psi_0 = e^A \sket{0_{\h}^+}\otimes \sket{0_{\h}^-}\quad, \eqno(3.23)$$
where $\sket{0_\h} = \sket{0_\h^+}\otimes\sket{0_\h^-}$ is the vacuum for
the coordinates (3.20), while
$$A = \int_0^{\infty}dk [f_0 (k) a_{k}^{\dagger} - f_0^{*}(k) a_{k}]
\quad.\eqno(3.24)$$
$f_0 (k)$ are the Fourier modes of $f_0 (x^+)$. Note that the operator
$A$ is written as a
linear combination of Kruskal coordinates creation and anhilation
operators, which is different from taking a corresponding linear
combination of the ``in'' operators, what was done in
\cite{mik4}. This choice may look unnatural, but it simplifies the
calculations and it gives the result of Bose, Parker and Peleg.

\sect{4. Semiclassical metric}

Now we are ready to calculate the effective metric (3.13). The lowest
order semiclassical metric will be given by the eq. (3.15). This
requires calculating the expectation values of the $T_{\pm\pm}$
operators, and a useful formula is
$$ \sbra{0_\h } T(x) \sket{0_\h} = -{1\over 48\p}\left(
{\h^{\prime\prime\prime} \over\h^{\prime}} -
\frac32 \left({\h^{\prime\prime}\over\h^{\prime}}\right)^2\right)
= -{1\over 48\p}D_x (\h)\quad.\eqno(4.1) $$

In the left-moving sector we have
$$\li{\sbra{0_\h }e^{-A} T_{++} e^A \sket{0_\h} =&
\sbra{0_\h } T_{++} - [A,T_{++}] + \ha [A,[A,T_{++}]] \sket{0_\h}\cr =&
 -{\k\over 4 (x^+)^2} + \ha \left({\pa f_0\over \pa x^+}\right)^2
\quad,&(4.2)\cr}$$
where $\k = {N\over 24 \p}$. In the right-moving sector we obtain
$$\sbra{0_\h } T_{--} \sket{0_\h} =  -{\k\over 4 (x^-)^2} \eqno(4.3)$$
from the formula (4.1).
By inserting the eqs (4.2-3) into the formula (3.15) we obtain
$$e^{-\r_0} = C + b_{\pm}x^\pm - \l^2 x^+ x^-
- {\k\over 4} \log (-\l^2 x^+ x^- ) -
 \ha\int_{\L^+}^{x^+} dy^+ (x^+ - y^+)\left({\pa f_0\over \pa
y^+}\right)^2 \eqno(4.4)$$
which is the BPP solution. It is a solution of the equations of motion of
an effective one-loop action
$$S_{eff} = S_0 - {N\over 96\p}\int \sqrt{-g}R\bo^{-1}R -
{N\over 24\p} \int \sqrt{-g}(R \f -(\nabla \f )^2 ) \eqno(4.5)$$
where $S_0$ is the classical CGHS action \cite{bpp}.

By setting $b_\pm =0$ and by choosing $C=\frac14 \k [1 - \log (\k/4)] $
one can obtain a
consistent  semiclassical geometry \cite{bpp}. In the case of
the shock-wave matter, one has a static geometry for $x^+ < x_0^+$
$$e^{-\r_0}= e^{-\f_0} = C - \l^2 x^+ x^-
- {\k\over 4} \log (-\l^2 x^+ x^- ) \eqno(4.6)$$
which is defined for $\s < \s_{cr}$, where $\s$ is the spatial
coordinate. At $\s = \s_{cr}$ there is a singularity, and this line is
interpreted as a boundary of a strong coupling region. The same phenomenon
occurs for the RST metric \cite{rst}, and a consistent geometry can be defined
for $\s < \s_{cr}$ by imposing a refelecting boundary conditions at
$\s = \s_{cr}$.

For $x^+ > x_0^+$ one obtains an evaporating black hole solution
$$e^{-\r_0} = C + {M\over \l} - \l^2 x^+ (x^- + \D )
- {\k\over 4} \log (-\l^2 x^+ x^- ) \quad.\eqno(4.7)$$
The corresponding Hawking radiation flux at future null-infinity is
determined by evaluating
$$ \sbra{0_\h} T_{--} (\x^-) \sket{0_\h} \eqno(4.8)$$
where $T_{--}(\x^-)$ is  normal ordered with respect to
asymptoticaly flat coordinates $\x^{\pm}$ at future null-infinity of
the metric (4.7). The $\x^{\pm}$ coordinates are the same as the out
coordinates (3.21) of the classical black hole solution. Then by using (4.1)
one obtains
$$ 2\p \svev{T_{--}(\s^-)} = {\l^2\over 48}\left[ 1 - (1 +
\l\D e^{\l\s^-})^{-2}\right]\quad,\eqno(4.9)$$
which corresponds to the thermal Hawking radiation, with $T_H = {\l\over 2\p}$
\cite{{cghs},{gn}}. The Hawking radiation shrinks the apparent horizon
of the solution (4.7), so that the apparent horizon line meets the
curvature singularity in a finite proper time, at $(x_{int}^+, x_{int}^-)$,
after which the
singularity becomes naked. However, a static solution
$$e^{-\r_0} = C  - \l^2 x^+ (x^- + \D )
- {\k\over 4} \log (-\l^2 x^+ (x^- + \D) )\eqno(4.10)$$
can be continuously matched to (4.7) along $x^- = x_{int}^-$. A small
negative energy shock-wave emanates from that point, and for $x^- >
x_{int}^-$ the Hawking radiation stops, and the static geometry (4.10) has
a null ADM mass. There is again a critical line $\s = \s_{cr}$,
corresponding to a singularity of the metric (4.10), which can be
interpreted as the boundary of the region where higher order
corrections become important. The spatial geometry of the remnant (4.10)
is that of a semi-infinite throat, extending to the strong coupling region.

Given the formula (3.13) we have a way to see when the BPP geometry is a
good approximation. This will happen if
$$ e^{2\r_0} \svev{\d F^2} << 1 \quad.\eqno(4.11)$$
This condition will generically fail for $e^{-\r_0} =0$, i.e. at the
singularities of the BPP metric.

\sect{5. Second order corrections}

Calculating the $\svev{\d F^n}$ terms will require calculating
$ \svev{ T(x_1) \cdots T(x_2) } $ and this will require a regularization.
As was discussed in \cite{mik4},
the singularity structure of such an expression is encoded in the
operator product expansion
$$ T (x) T (y) = {c/2\over (x - y)^4} +
{2T (y)\over (x - y)^2} + {2\pa T (y)\over x - y }
+ {\rm const.} + o(x - y)\quad, \eqno(5.1)$$
The simplest ordering (regularization) is
$$:T(x_1) \cdots T(x_n): = T(x_1) \cdots T(x_n) -\sbra{0}T(x_1) \cdots T(x_n)
\sket{0} \quad,\eqno(5.2)$$
where $\sket{0}$ is the relevant vacuum (i.e. $\sbra{0}T\sket{0}
=0$). However, this removes only the leading divergencies coming from the
$(x-y)^{-4}$ term, and more sophisticated schemes can be employed
based on the point-splitting method \cite{mik4}. We apply these
methods for calculating the second order corrections.

It will be usefull to redefine $F$ as
$$\li{e^{-\f} =&  \a + \b_{\pm}x^\pm -\l^2 x^+ x^-
- \int_{\L^\pm}^{x^\pm} dy (x - y)\sbra{0_\h} T_{\pm\pm}(y)\sket{0_\h} \cr -&
 \ha\int_{\L^\pm}^{x^\pm} dy^+ (x^+ - y^+)\left(T_{\pm\pm}(y) - \sbra{0_\h}
T_{\pm\pm}(y) \sket{0_\h} \right)\cr
=& C -\l^2 x^+ x^-  - {\k\over 4} \log (-\l^2 x^+ x^- ) - F_+ -F_- &(5.3)\cr}$$
so that new $F_{\pm}$ are given as
$$F_{\pm} = \int_{\L^{\pm}}^{x^{\pm}} dy (x^{\pm} - y) \tilde{T}_{\pm\pm} (y)
 \eqno(5.4)$$
where
$ \tilde{T} = T - \sbra{0_\h} T \sket{0_\h} $
and $\sbra{0_\h} \tilde{T} \sket{0_\h} = 0$. Another convinient
redefinition is to rescale $f(x)$ to $\sqrt{2\p}f(x)$.

In the left sector we define
$$:T(x_1) T(x_2): = T(x_1) T(x_2) -\sbra{0_\h}T(x_1) T(x_2)
\sket{0_\h} \quad,\eqno(5.5)$$
so that
$$\li{\sbra{0_\h}:\tilde{T}_A (x_1) \tilde{T}_A (x_2): \sket{0_\h}
=& \sbra{0_\h} [A,T_1] [A,T_2]\sket{0_\h}\cr
=&-{1\over 4\p^2}{\pa f_0\over\pa x_1}{\pa f_0\over\pa x_2}
\sbra{0_\h}{\pa f\over\pa x_1}{\pa f\over\pa x_2}\sket{0_\h}\cr
=& {1\over 8\p^2} {\pa f_0\over\pa x_1}{\pa f_0\over\pa x_2}
\pa_{x_1} \pa_{x_2} \log |\h (x_1) - \h(x_2)|\quad.&(5.6)\cr}$$
where we have used
$$  \sbra{0_{\h}} f (x_1) f (x_2) \sket{0_{\h}}
= -\ha \log |\h (x_1) - \h (x_2 )|  \quad.\eqno(5.7)$$
We have omitted the $+$ indicies for the simplicity sake.

The expression (5.6) is still divergent when $x_1 \to x_2$ since
$$\pa_{x_1} \pa_{x_2} \log |\h (x_1) - \h(x_2)| =
{1\over(x_1 - x_2)^2} + \frac16 D_{x_1}(\h) + o(x_1 - x_2) \quad.
\eqno(5.8)$$
However, one can use
$$ \int_0^{\infty} dk\,k\, e^{ik(\h_1 - \h_2 )} =
-{1\over (\h_1 - \h_2 )^2} \quad,\quad\h_1 - \h_2 \ne 0 \quad,\eqno(5.9)$$
to rewrite (5.6) as
$$-{1\over 8\p^2}\int_0^{\infty} dk\,k\, e^{ik(\h_1 - \h_2 )}
\pa_{x_1}\h\pa_{x_2}\h \pa_{x_1} f_0 \pa_{x_2} f_0 \quad.\eqno(5.10)$$
This gives
$$\li{\svev{\d F_+^2} =& -{1\over 8\p^2 \l^4} \prod_{i=1}^2
\int_{-\infty}^{\h^+} d\h_i
(e^{\l(\h^+ - \h_i)} - 1) \int_0^{\infty} dk\,k\, e^{ik(\h_1 -\h_2 )}
{\pa f_0\over\pa \h_1}{\pa f_0\over\pa \h_2}\cr =&
-{1\over 8\p^2 \l^4}\int_0^{\infty} dk\,k\, |\cf (k, \h^+)|^2
\quad,&(5.11)\cr}$$
where
$$\cf (k,\h^+) = \int_{-\infty}^{\h^+} d\h e^{ik\h}(e^{\l(\h^+ - \h)} - 1)
{\pa f_0\over\pa \h} \quad.\eqno(5.12)$$
Then by choosing $f_0$ which falls quickly enough away from the centre of the
matter pulse, we can get $\cf (k)$ such that (5.11) is
finite. However, if one wants convergence for arbitrary $f_0$'s  of compact
support, then additional regularization is needed. One way would be by
subtracting $(x_1 - x_2)^{-2}$ from $\pa_1 \pa_2 \log(\h_1 - \h_2)$,
which corresponds to using a new ordering
$$\li{:T(x_1) T(x_2): =& T(x_1) T(x_2) -\sbra{0_\h}T(x_1) T(x_2)
\sket{0_\h}\cr -& \sbra{0_x}[A,T(x_1)][A,T(x_2)]\sket{0_x}  \quad.&(5.12)\cr}$$
This gives for $\svev{\d F_+^2}$
$$ {1\over 8\p^2} \prod_{i=1}^2 \int_{\L^+}^{x^+}
dx_i  (x^+ - x_i )\left( \pa_{x_1} \pa_{x_2}
\log |\h^+ (x_1) - \h^+ (x_2)|- {1\over (x_1 - x_2 )^2} \right)
{\pa f_0\over\pa x_1}{\pa f_0\over\pa x_2}\,.
 \eqno(5.13)$$

In the right sector we use the point-splitting method for regularizing the
operator products. It amounts to calculating an appropriate
limes ($x_{2i-1} \to x_{2i}$) of the expression
$$ \sbra{0_{\h}} \prod_{i=1}^{2n} \pa_{x_i^-} f (\h (x_i)) \sket{0_{\h}}
\quad.\eqno(5.14)$$
Expression (5.13) can be calculated by using Wick's theorem and by using
the expression for a two-point function (5.7).
A particular normal ordering can be chosen by an appropriate subtraction of
the products of the terms $(x_i^- - x_j^-)^{-2}$ and
$\pa_k f_k\cdots \pa_l f_l $ from the expression (5.13)
before taking the limes, such that one obtains a regular expression
after taking the limes. A useful formula for doing this is (5.8).

For example,
$$2\p T(x_1) = \lim_{x_2 \to x_1}\ha \left( {\pa f\over\pa x_1}{\pa f\over\pa
x_2} +\ha {1\over(x_1 - x_2)^2} \right) \eqno(5.15)$$
which gives the result (4.1). In the $n=2$ case, one can define analogously
$$\li{4\p^2 :T(x_1) T(x_2): = &\lim_{x_3 \to x_1}\lim_{x_4 \to x_2}
\frac14 \Big(  {\pa f\over\pa x_1}{\pa f\over\pa x_3}{\pa f\over\pa x_2}
{\pa f\over\pa x_4} +\ha
{1\over(x_1 - x_3)^2}{\pa f\over\pa x_2}{\pa f\over\pa x_4}\cr
&+\ha {1\over(x_2 - x_4)^2}{\pa f\over\pa x_1}{\pa f\over\pa x_3}
+ \frac14 {1\over(x_1 - x_3)^2}{1\over(x_2 - x_4)^2}
 \Big) \quad.&(5.16)\cr}$$
This gives
$$\li{ \sbra{0_{\h}}: T_{--}(x_1) T_{--} (x_2): \sket{0_{\h}}
=&{1\over 32\p^2}\left( \pa_{x_1} \pa_{x_2}
\log |\h^- (x_1) - \h^- (x_2)|  \right)^2\cr
+&\svev{T_{--} (x_1)}\svev{T_{--} (x_2)}\quad.&(5.17)\cr}$$
The expression (5.17) is still divergent when $x_1 \to x_2$, and it will be
the source of divergence in
$$\svev{\d F_-^2}= {1\over 32\p^2}\int_{\L^-}^{x^-}\int_{\L^-}^{x^-}
dx_1 dx_2  (x^- - x_1 )(x^- - x_2 )\left( \pa_{x_1} \pa_{x_2}
\log |\h^- (x_1) - \h^- (x_2)| \right)^2 \, .
\eqno(5.18)$$
One way to regularize (5.18) is by changing the definition (5.16)
\cite{mik4}. One can define
$$\li{4\p^2 & : T(x_1) T(x_2): = \lim_{x_3 \to x_1}\lim_{x_4 \to x_2}
\Big(\frac14 {\pa f\over\pa x_1}{\pa f\over\pa x_3}{\pa f\over\pa x_2}
{\pa f\over\pa x_4}\cr +
&\ha{1\over(x_1 - x_3)^2}{\pa f\over\pa x_2}{\pa f\over\pa x_4}
+\ha {1\over(x_2 - x_4)^2}{\pa f\over\pa x_1}{\pa f\over\pa x_3}
+\frac14 {1\over(x_1 - x_3)^2}{1\over(x_2 - x_4)^2}\cr
+&\ha{1\over(x_1 - x_2)^2}{\pa f\over\pa x_3}{\pa f\over\pa x_4}
+\ha {1\over(x_3 - x_4)^2}{\pa f\over\pa x_1}{\pa f\over\pa x_2}
+\frac14 {1\over(x_1 - x_2)^2}{1\over(x_3 - x_4)^2}\cr
+&\ha{1\over(x_1 - x_4)^2}{\pa f\over\pa x_2}{\pa f\over\pa x_3}
+\ha {1\over(x_2 - x_3)^2}{\pa f\over\pa x_1}{\pa f\over\pa x_4}
+\frac14 {1\over(x_1 - x_4)^2}{1\over(x_2 - x_3)^2}\Big) \, ,\cr
&\quad &(5.19)\cr}$$
so that
$$\li{ \sbra{0_{\h}}: T_{--}(x_1) T_{--} (x_2): \sket{0_{\h}}
=&{1\over 32\p^2}\left( \pa_{x_1} \pa_{x_2} \log |\h^- (x_1) - \h^- (x_2)|
- {1\over (x_1 - x_2)^2}\right)^2\cr
+&\svev{T_{--} (x_1)}\svev{T_{--} (x_2)}\quad.&(5.20)\cr}$$
(5.20) is finite for $x_1 \to x_2$ due to eq. (5.8). Consequently
$\svev{\d F_-^2}$ is given by
$$ {1\over 32\p^2} \prod_{i=1}^2 \int_{\L^-}^{x^-}
dx_i  (x^- - x_i )\left( \pa_{x_1} \pa_{x_2}
\log |\h^- (x_1) - \h^- (x_2)|- {1\over (x_1 - x_2 )^2} \right)^2\,,
\eqno(5.21)$$
which is finite.

\sect{6. Conclussions}

We have shown that in a unitary gauge quantization of the CGHS model one
can obtain an evaporating black hole solution in the semiclassical
limit, which is also a solution of the one-loop effective action
proposed by BPP, which is manifestly diffeomorphism invariant.
This partially answers a dillema about the
diffeomorphism invariance of our construction, since it means that the
diffeomorphism invariance is preserved in the lowest order of the
expansion (3.13). However, the drawback is that it is not obvious
which regularization procedure will give a diffeomorphism invariant expressions
for the higher order corrections $\svev{\d F^n }$, ($n\ge 2$). This
problem is being investigated \cite{mr}.

Also note that in the unitary gauge quantization there is no restriction on
the number of matter fields $N$, and one could take $N=1$. Since the
constraints of the theory can be transformed into the constraints of an $(N+2)$
-dimensional bosonic string \cite{mik1,hks,mik3,cj},
one could worry about the diffeo
anomalies for $N \ne 24$, which appear in the Dirac quantization of
the bosonic string. However, Kuchar and Torre have shown that an
anomaly free Dirac quantization of the bosonic string is possible for
any $N$ \cite{kt}.
Although the Kuchar-Torre construction requires canonical variables
such that the constraints become non-polynomial, and that makes them
difficult to solve, it shows that it is possible to preserve the
diffeo invariance for $N \ne 24$. Hence it is not surprising that in a
unitary gauge quantization
approach the diffeomorphism invariance is preserved at one-loop order.
This also lays at rest doubts that a diffeomorphism invariant
quantization of the CGHS model can give a free-field Fock space as a
physical Hilbert space \cite{cjz}.

The quantum gravity theory we have constructed has a fixed space-time manifold
${\bf R}^2$, with differential structure (i.e. coordinates) but without metric.
The wavefunctions of the
theory are defined on $t=$const. surfaces, which are related by a
unitary transformation (3.7). The dilaton and the metric are induced
by the matter, which propagates as a free field. The space-time
geometry is given by the expectation value of the metric and the dilaton
operator, and it is defined only in the regions of ${\bf R}^2$ where the
quantum fluctuations are small. The one-loop results indicate that the
quantum fluctuations become large at the curvature singularity, which
would indicate a notion of a classical singularity as a place where
the geometry becomes fuzzy.

The unitarity of the theory implies that an observer located at
$x =\infty$ who sends the matter pulse to $x= -\infty$ at $t= -\infty$
will generically know the complete wavefunction at later times.
However, when observer performs a measurement, the wavefunction
will collapse into an eigenvalue
state of his observables. If the measurement is performed before the
time $t_0$ when the apparent horizon forms in the effective metric (3.10),
the outcome of the
measurement will be completely known to the observer, and it will be a
state in $\ch$. For
$t>t_0$ the outcome of the measurement will be again a state in $\ch$,
but the observer
will only know the part of the state which is associated to the
observables lying outside of the horizon. Hence for $t>t_0$ he will
have to use a density matrix in
the Hilbert space $\ch_{outs}(t)$, to describe the results of his
measurements ($\ch_{outs}$ is the Hilbert space corresponding to the
observables defined outside of the horizon).
However, because of the unitarity, the relation
$$ \ch = \ch_{ins}(t) \otimes \ch_{outs}(t)$$
will be always valid ($\ch_{ins}$ is the Hilbert space corresponding
to observables inside the horizon). Hence whether the observer will
see a pure state or a mixed state will depend on the behaviour of the
exact quantum metric (3.10). Given the one-loop approximation (3.15),
it is still unclear what will be the effect of the higher order
corrections in the strong copupling region, but since the theory is
unitary, one has two known possibilities: remnants or the complete
evaporation with the information being returned through the radiation.
The one-loop result seems to favour the remnant scenario.
However, a further work is neccessary for an answer to this question.

\sect{Acknowledgements}

\noindent I would like to thank Voja Radovanovi\'c and Don Marolf for useful
discussions.

\end{document}